\documentclass[letterpaper, 11 pt, conference, onecolumn]{article}

\usepackage{graphics} 
\usepackage{epsfig} 
\usepackage{mathptmx} 
\usepackage{times} 
\usepackage{amsmath} 
\usepackage{amssymb}  
\usepackage{hyperref}
\usepackage[font=small]{caption}
\usepackage{pdfpages}
\usepackage{algorithm}
\usepackage[noend]{algpseudocode}

\title{\LARGE \bf
Modeling Baseball Outcomes as Higher-Order Markov Chains
}
\author{Jun Hee Kim \\ \texttt{\small junheek1@andrew.cmu.edu} \\  \textit{Department of Statistics \& Data Science, Carnegie Mellon University}}
\date{}

\begin{document}

\maketitle
\thispagestyle{plain}
\pagestyle{plain}
\footskip = 30pt
\hoffset = 0pt
\paperwidth = 597pt

\begin{abstract}
\normalfont Baseball is one of the few sports in which each team plays a game nearly everyday. For instance, in the baseball league in South Korea, namely the KBO (Korea Baseball Organization) league, every team has a game everyday except for Mondays. This consecutiveness of the KBO league schedule could make a team's match outcome be associated to the results of recent games. This paper deals with modeling the match outcomes of each of the ten teams in the KBO league as a higher-order Markov chain, where the possible states are win ($``W"$), draw ($``D"$), and loss ($``L"$). For each team, the value of $k$ in which the $k^{\text{th}}$ order Markov chain model best describes the match outcome sequence is computed. Further, whether there are any patterns between such a value of k and the team's overall performance in the league is examined. We find that for the top three teams in the league, lower values of $k$ tend to have the $k^{th}$ order Markov chain to better model their outcome, but the other teams don't reveal such patterns.
\end{abstract}

\section{Introduction}
\noindent
Different types of sports have different game rules, frequency of games, style, etc that make the sport unique and interesting. In particular, how often games are held is a factor that differs a lot across different sports. Baseball, which is a sport that the league is very popular in countries like United States (Major League Baseball, or MLB), South Korea (Korea Baseball Organization league, or KBO league), and Japan (Nippon Professional Baseball, or NPB), has a feature that each team plays a game nearly everyday. This is in contrast to, say, the English Premier League (the soccer league in England), where each team has a game once or twice a week. In particular, in the KBO league, each of the ten teams has a match everyday except for Mondays. This feature that there is a game nearly everyday could be associated somehow to a team's performance in a game. For instance, a team might be more likely to do well on a game if it had won all the recent five games compared to the case where it had lost all the recent three games. In fact, when discussing a preview or prediction on a game, many KBO league news articles mention the number of consecutive wins or losses the team currently has.

This paper focuses on the KBO league, in particular how the consecutiveness of the schedule is related to each team's outcome. More specifically, this paper examines, for each of the ten teams in the KBO league, if we were to model the game outcomes (represented as a single sequence) as a $k^{\text{th}}$ order Markov chain, which value of $k$ is the most effective. In section 2, we introduce the KBO league in general, and in section 3, the higher-order Markov chain model whose possible states are win ($``W"$), draw ($``D"$), and loss ($``L"$), particularly the one used in the \textsf{markovchain} R package$^{4}$, is discussed. Then in sections 4 and 5, how we assess the model fit and the actual model results are reported, and lastly in section 6 we discuss conclusions and potential future work.

\section{KBO League Introduction}
\noindent
In this section, we introduce the KBO league in general. The KBO league began in 1982 with six teams back then: Haitai Tigers, Lotte Giants, MBC Blue Dragons, OB Bears, Sammi Superstars, and Samsung Lions$^{1}$. With some teams being the successor of the teams back then in 1982, currently the KBO league has ten teams competing: Doosan Bears, SK Wyverns, Hanwha Eagles, Nexen Heroes, LG Twins, Samsung Lions, Lotte Giants, KIA Tigers, KT Wiz, and NC Dinos. Unlike in the MLB (where team names represent the home location: e.g. New York Yankees), the KBO league team names represent the sponsor corporation. Furthermore, unlike in MLB where the game doesn't end as a draw (or a tie) except for exceptional reasons like weather (or sometimes darkenss), in a KBO league game, if the two teams have the same score after the 12th inning, the game ends as a draw.

The league does not have sub-leagues. Rather, the ten teams together compete in the pennant race in such a way that each of the ten teams face the other nine teams 16 times, eight home games and eight away games, thereby playing total 144 games in the regular season. In post-season, the 5th place team competes in a wild-card round against the 4th place team. In the wild-card round, if the 4th place wins the first game, then the round is immediately over with the 4th place going to the semi-playoffs, but if the 5th place wins the first game, then the two teams compete in the second game where that game's winner goes to the semi-playoffs. The wild-card round victor faces the 3rd place team in the semi-playoffs with a best-of-five format, and the semi-playoffs victor faces the 2nd place in the playoffs, also in best-of-five. Finally, the playoffs victor plays against the 1st place team in the final round called the Korean Series with a best-of-seven format. Note that the rules mentioned in this paragraph could change in future seasons (for example, in the 2015 season the total number of games changed from 128 to 144), but at least in the 2018 season, those rules are applied$^{2}$.

Table 1 shows the ranking of the KBO league 2018 as of August 18th, 2018, which is right before the Asian Games break (the KBO league in 2018 has a break of approximately 3 weeks since some of the players go to the Jakarta-Palembang Asian Games 2018 as part of the South Korean national team).

\begin{table}[htbp]
\begin{center}
\begin{tabular}{|c|c|c|c|c|c|c|c|}
\hline
Rank & Team & Games & Wins & Draws & Losses & Winning rate & Games behind \\ \hline
1 & Doosan Bears & 113 & 73 & 0 & 40 & 0.646 & 0.0 \\ \hline
2 & SK Wyverns & 112 & 62 & 1 & 49 & 0.559 & 10.0 \\ \hline
3 & Hanwha Eagles & 114 & 62 & 0 & 52 & 0.544 & 11.5 \\ \hline
4 & Nexen Heroes & 118 & 61 & 0 & 57 & 0.517 & 14.5 \\ \hline
5 & LG Twins & 116 & 56 & 1 & 59 & 0.487 & 18.0 \\ \hline
6 & Samsung Lions & 116 & 54 & 3 & 59 & 0.478 & 19.0 \\ \hline
7 & Lotte Giants & 110 & 51 & 2 & 57 & 0.472 & 19.5 \\ \hline
8 & KIA Tigers & 110 & 51 & 0 & 59 & 0.464 & 20.5 \\ \hline
9 & KT Wiz & 113 & 47 & 2 & 64 & 0.423 & 25.0 \\ \hline
10 & NC Dinos & 116 & 47 & 1 & 68 & 0.409 & 27.0 \\ \hline
\end{tabular}
	\caption{KBO League 2018 Rank (as of August 18th, 2018)}
	\label{tab:num1}
\end{center}
\end{table}

\section{Higher-Order Markov Chain Model}
\noindent
The goal of this paper is to use higher-order Markov chains to model the game outcomes for each team in the KBO league. This section introduces the higher-order Markov chain model and parameter estimation methods. The notations and formulations of the model discussed in this section follow Chapter 6 in Ching, Huang, Ng, and Siu (2013)$^{3}$, in which the \textsf{markovchain} R package$^{4}$, which we use for the computation in this study, was implemented based on.

\subsection{First-order Markov Chain}

First, we briefly introduce the (discrete time) first-order Markov chain, usually referred to just ``Markov chain". Let the data sequence $\{ X^{(i)} \}_{i=1}^{n} = \{ X^{(1)}, X^{(2)}, \cdots, X^{(n)} \} $ be a stochastic process where each $X^{(i)}$ can take finite or countable number of possible values. We call such possible values `states'. For example, a stochastic process whose possible states are ``sunny'', ``cloudy'', and ``rainy'' may produce a data sequence of $\{ \text{``cloudy''}, \text{``cloudy''}, \text{``rainy''}, \text{``sunny''}, \text{``rainy''}  \}$. The set of all possible states is called the `state space', which can consist of essentially anything: numbers, letters, weather conditions, baseball game outcomes, etc. In this paper, we let $S$ denote the state space, and let $m$ denote the number of possible states (i.e. $m = |S|$).

The name ``Markov chain" comes from the (first-order) Markov property. The property states, or assumes, that the next state $X^{(n)}$ is conditionally independent of all the states so far (i.e. $X^{(n-1)}, X^{(n-2)}, \cdots, X^{(1)}$) given the current state $X^{(n-1)}$. That is: for any timestep $n$,

\begin{equation}
P(X^{(n)} = x_{new} | X^{(n-1)}=x_{n-1}, X^{(n-2)}=x_{n-2}, \cdots, X^{(1)}=x_{1}) = P(X^{(n)} = x_{new} | X^{(n-1)}=x_{n-1})
\end{equation}
Intuitively, we wander around the states in the state space, and the most recent past is only what matters for the present.

Moreover, the model assumes that for each pair of states $(i, j)$, there is a fixed transition probability $p_{ij}$, which is the probability that the process moves to state $i$ given that it's currently at state $j$. The chain always decides its state at the next timestep according to these transition probabilities, which can be represented as a single $m \times m$ matrix called the ``transition matrix". In our notation, the row $i$ \& column $j$ entry of the transition matrix has $p_{ij}$, the transition probability from state $j$ to state $i$. Intuitively, we can think of each column of the transition matrix representing the ``from" state, and each row being the ``to" state. Clearly, each column of the transition matrix must sum to 1.

\subsection{Higher-order Markov Chain}

In the first-order Markov chain model, the assumption was that the state at timestep $n$ only depends on the state at the timestep immediately before (i.e. $n-1$) and all the further past are meaningless. We can relax the assumption in such a way that the state at a timestep depends on more of the recent past. Formally, a $k^{th}$ order Markov chain assumes that the state at timestep $n$ only depends on the states at the recent $k$ timesteps (i.e. $n-1, n-2, \cdots, n-k$). That is: for any timestep $n$:

\begin{equation}
\begin{split}
& \quad \enskip P(X^{(n)} = x_{new} | X^{(n-1)}=x_{n-1}, X^{(n-2)}=x_{n-2}, \cdots, X^{(1)}=x_{1}) \\
&= P(X^{(n)} = x_{new} | X^{(n-1)}=x_{n-1}, X^{(n-2)}=x_{n-2}, \cdots, X^{(n-k)}=x_{n-k})
\end{split}
\end{equation}
Notice that if we set $k=1$, then the model is equivalent to what was introduced in Section 3.1, and this is why it is called the ``first-order Markov chain".

Furthermore, the $k^{th}$ order Markov chain model assumes that there is an $m \times m$ transition matrix $Q^{(l)}$ defined for each lag $l \in \{1, \cdots, k\}$. The row $i$ \& column $j$ entry of the $l$-step transition matrix $Q^{(l)}$ has the probability that the process will move to state $i$ after $l$ timesteps given that currently it's at state $j$. Again, clearly it must be true that each column of $Q^{(l)}$ sums to 1, $\forall l \in \{1, \cdots, k\}$. Also, each lag $l \in \{1, \cdots, k\}$ has a non-negative weight $\lambda_{l}$ with: 

\begin{equation}
\sum_{l=1}^{k} \lambda_{l} = 1
\end{equation}

Then, the model says:

\begin{equation}
\mathbf{X}^{(n+k+1)} = \sum_{l=1}^{k} \lambda_{l} Q^{(l)} \mathbf{X}^{(n+k+1-l)}
\end{equation}
where $\mathbf{X}^{(n+k+1-l)}$ is an $m \times 1$ vector that shows the probability distribution of the $m$ states at timestep $n+k+1-l$, which essentially shows, for each state $i$, if we draw this Markov chain process many times, what proportion of those simulations will be at state $i$ at timestep $n+k+1-l$.

Equation (4) can be rewritten as:

\begin{equation}
P(X^{(n)} = x_{new} | X^{(n-1)}=x_{n-1}, X^{(n-2)}=x_{n-2}, \cdots, X^{(n-k)}=x_{n-k}) = \sum_{l=1}^{k} \lambda_{l} q_{x_{new}, x_{n-l}}^{(l)}
\end{equation}
where $q_{x_{new}, x_{n-l}}^{(l)}$ denotes the row $x_{new}$ \& column $x_{n-1}$ entry of the matrix $	Q^{(l)} $. It can be shown that if $Q^{(l)}$ is irreducible and aperiodic, $\lambda_{l} > 0$, and $\sum_{l=1}^{k} \lambda_{l} = 1$, then this model has a stationary distribution $\mathbf{X}$ that satisfies $\Big( \mathbf{I} - \sum_{l=1}^{k} \lambda_{l} Q^{(l)} \Big) \mathbf{X} = \mathbf{0}$ and also $\text{lim}_{n \rightarrow \infty} \mathbf{X}^{(n)} = \mathbf{X}$, where $\mathbf{I}$ denotes the $m \times m$ identity matrix, and $\mathbf{0}$ is the length-$m$ vector of all $0$'s.

Now we discuss the methods for estimating the model parameters: $Q^{(l)}$ and $\lambda_{l}$ for each $\l \in \{1, \cdots, k\}$. Notice that this higher-order Markov chain model has $k + km^{2}$ parameters since each transition matrix $Q^{(l)}$ has $m^{2}$ entries.

Again, assume we observe a data sequence of length $n$: $\{ X^{(t)} \}_{t=1}^{n} = \{ X^{(1)}, X^{(2)}, \cdots, X^{(n)} \} $. For every ordered pair of states $(i,j)$, for each lag $l \in \{1, \cdots, k\}$, we define the transition frequency $f_{ji}^{(l)}$ as the number of times in the given data sequence such that the process is at state $i$ and then after $l$ steps it is at state $j$. Naturally, we can write these altogether in matrix form: we define the $l$-step transition frequency matrix $F^{(l)}$ (of size $m \times m$) as:

\begin{equation}
F^{(l)} = \begin{bmatrix}
f_{11}^{(l)} & f_{12}^{(l)} & \cdots & f_{1m}^{(l)} \\
f_{21}^{(l)} & f_{22}^{(l)} & \cdots & f_{2m}^{(l)} \\
\vdots & \vdots & \ddots & \vdots \\
f_{m1}^{(l)} & f_{m2}^{(l)} & \cdots & f_{mm}^{(l)} \\
\end{bmatrix}
\end{equation}
Of course, this matrix is defined for every lag $l \in \{1, \cdots, k\}$.

Then, for each lag $l \in \{1, \cdots, k\}$, we can estimate the $l$-step transition matrix $Q^{(l)}$ as:

\begin{equation}
\hat{Q}^{(l)} = \begin{bmatrix}
\hat{q}_{11}^{(l)} & \hat{q}_{12}^{(l)} & \cdots & \hat{q}_{1m}^{(l)} \\
\hat{q}_{21}^{(l)} & \hat{q}_{22}^{(l)} & \cdots & \hat{q}_{2m}^{(l)} \\
\vdots & \vdots & \ddots & \vdots \\
\hat{q}_{m1}^{(l)} & \hat{q}_{m2}^{(l)} & \cdots & \hat{q}_{mm}^{(l)} \\
\end{bmatrix}
\end{equation}
where

\begin{equation}
\hat{q}_{ij}^{(l)} = \left\{
        \begin{array}{ll}
            \frac{f_{ij}^{(l)}}{\sum_{i=1}^{m}f_{ij}^{(l)}} & \quad \text{if } \sum_{i=1}^{m}f_{ij}^{(l)} \neq 0 \\
            0 & \quad \text{otherwise}
        \end{array}
    \right.
\end{equation}
Note that $\hat{q}_{ij}^{(l)} = 0$ if there is no observation such that the process is at state $j$ and then after $l$ steps it is at some state, which happens when state $j$ appears only at the last $l$ timesteps of the observed data sequence.

Also, the stationary distribution $\mathbf{X}$ can be estimated from the observed data sequence as the proportion of the occurrence of each state in the sequence. That is: for each state $i$, our estimate of the corresponding entry in the stationary distribution is just the number of times state $i$ appears in our length-$n$ sequence divided by $n$. Let's denote such an estimate by $\hat{\mathbf{X}}$.

Given the estimated transition matrices $\hat{Q}^{(1)}, \cdots, \hat{Q}^{(k)}$ and the estimated stationary distribution $\hat{\mathbf{X}}$, we can estimate the $\lambda_{l}$ parameters via solving the following linear programming problem:

\begin{equation}
\underset{\lambda}{\text{min}} \sum_{i=1}^{m} w_{i}
\end{equation}
\text{subject to}

\begin{equation}
\begin{bmatrix}
w_{1} \\ w_{2} \\ \vdots \\ w_{m}
\end{bmatrix}
\ge \hat{\mathbf{X}} - \Big[ \hat{Q}^{(1)} \hat{\mathbf{X}} \enskip | \enskip \hat{Q}^{(2)} \hat{\mathbf{X}} \enskip | \cdots | \enskip \hat{Q}^{(k)} \hat{\mathbf{X}} \Big] \begin{bmatrix}
\lambda_{1} \\ \lambda_{2} \\ \vdots \\ \lambda_{m}
\end{bmatrix} ,
\end{equation} 
\begin{equation}
\begin{bmatrix}
w_{1} \\ w_{2} \\ \vdots \\ w_{m}
\end{bmatrix}
\ge - \hat{\mathbf{X}} + \Big[ \hat{Q}^{(1)} \hat{\mathbf{X}} \enskip | \enskip \hat{Q}^{(2)} \hat{\mathbf{X}} \enskip | \cdots | \enskip \hat{Q}^{(k)} \hat{\mathbf{X}} \Big] \begin{bmatrix}
\lambda_{1} \\ \lambda_{2} \\ \vdots \\ \lambda_{m}
\end{bmatrix} ,
\end{equation}
\begin{equation}
\forall i \in \{ 1, \cdots, m \}. \enskip w_{i} \ge 0,
\end{equation}
\begin{equation}
\forall l \in \{ 1, \cdots, k \}. \lambda_{l} \ge 0,
\end{equation}
\begin{equation}
\sum_{l=1}^{k} \lambda_{l} = 1
\end{equation}

\section{Method for Assessing Model Fit}
\noindent
Now that we know how the model is defined and how the parameters are estimated (in the \textsf{markovchain} R package$^{4}$), in this section, we introduce how we assess the quality of the model fit, given a fitted higher-order Markov chain model.

For each of the ten teams in the KBO league, we fit a $k^{th}$ order Markov chain on its data sequence of the outcomes of the recent 100 games, for $k = 1, \cdots, 13$. Here, the state space is $\{ ``W", ``D", ``L" \}$ where each state (in the listed order) represents win, draw, and loss, respectively. Each fitted object in the \textsf{markovchain} R package$^{4}$ returns the estimated $\lambda_{l}$ parameters, the estimated $Q^{(l)}$ matrices, and the estimated stationary distribution $\mathbf{X}$. We assess which value of $k$ has the corresponding $k^{th}$ order Markov chain model best describing the team's data sequence via the following procedure. For each team:

\noindent\rule{8cm}{0.4pt}

\begin{algorithmic}
\State $tenGames \gets \text{Randomly choose 10 out of the 100 games in the team's data sequence}$
\For {$k \text{ in } \{1, \cdots, 13 \}$}
\For {$game \text{ in } tenGames$}
\For {$state \text{ in } \{ ``W", ``D", ``L" \}$}
\State $p_{state} \gets P(game=state | \text{recent $k$ observations})$ computed via Equation (5) \enskip (We'll get $p_{W}, p_{D}, p_{L}$)
\EndFor
\State $predict \gets X \sim Categorical(p_{W}, p_{D}, p_{L})$
\EndFor
\EndFor
\State $team\_k\_acc \gets (\text{number of correct predictions}) / 10$
\end{algorithmic}

\noindent\rule{8cm}{0.4pt}

In words, for each team, we first randomly select 10 games out of the 100 present in the team's sequence. We examine across every value of $k$ (corresponding to the $k^{th}$ order Markov chain fitted to this team's sequence) via: 

\begin{enumerate}
\item For each of the ten games, for each of the three possible states, compute the estimated probability that the game's outcome was that particular state given the recent $k$ observations, using Equation (5) and the estimated $\lambda_{l}$'s and the $Q^{(l)}$'s.
\item Then, run a simulation from a Categorical distribution (which is essentially a generalization of the Bernoulli distribution where there can be more than two categories) that has three categories (``W'', ``D'', and ``L'') with the computed probabilities. The sampled outcome is our prediction on this game's result. Compare our prediction with the actual game outcome in the team's data sequence.
\item Calculate the prediction accuracy: Out of the ten predictions, how many are correct?
\end{enumerate}

After this process, for each team, we have the prediction accuracy of each of the 13 values of $k$. We assess the fit of the $k^{th}$ order Markov chain model applied to this team's sequence via how high the prediction accuracy is. That is: for each team, we rank the 13 values of $k$ on how well the $k^{th}$ order Markov chain modeled, or described, the observed length-$100$ sequence of the team.

\section{Results}
\noindent
Here we present the model fit results. For each team, We execute the process described in Section 4 and draw a barplot where each bar in the vertical axis represents each $k$ value, and the horizontal axis, of course, shows the prediction accuracy of the corresponding $k^{th}$ order Markov chain model fitted on that team's sequence. The barplots are shown in Figure 1.

\begin{figure}[h!]
\centering
  \includegraphics[width=16cm, height=20cm]{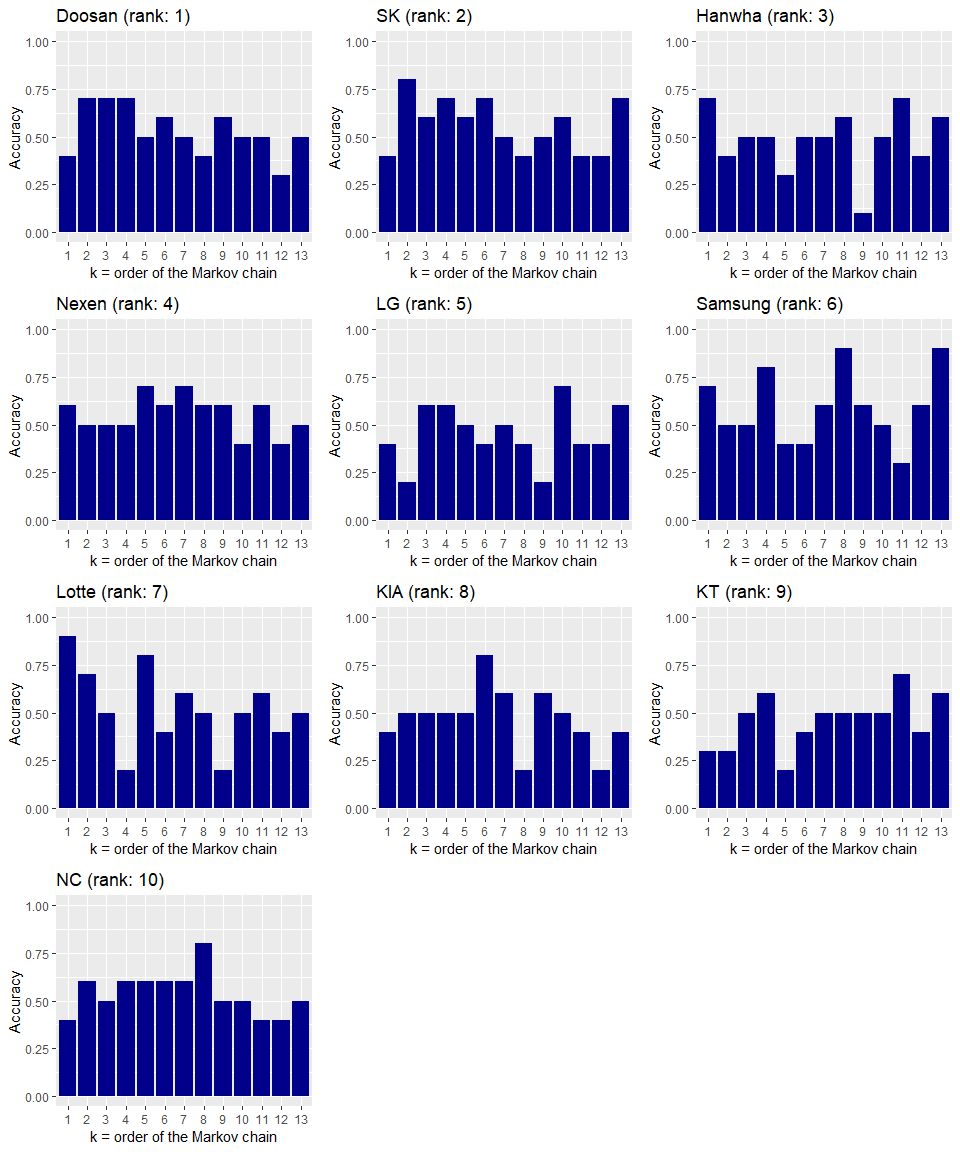}
  \caption{Model Fit Result for Each Team}
\end{figure}

Intuitively, if a particular value of $k$ has the highest prediction accuracy compared to the other values, we can think of it as: for predicting this team's performance of an arbitrary day, considering exactly the recent $k$ games' outcomes works the best, compared to any smaller or larger values of $k$. 

One of the most interesting patterns we can see in the plots that both Doosan Bears and SK Wyverns, which are the 1st and 2nd ranked teams in the league, have a skewed-to-the right shape (except that the lowest value $k=1$ has a low accuracy), meaning that lower values of $k$ tend to predict better than higher values. In particular, they both have $k = 2$ having the highest accuracy. This means that taking only the recent two matches into account best describes the team's performance in general. The 3rd ranked team Hanwha Eagles also has a right-skewed shape except that the accuracy rises again for the $k < 9$. So overall, the top 3 teams in the league tend to have their recent few games (say, one, two, and three) associated with the performance in a new game the most. Considering the fact that these teams all have relatively high winning rates, we could interpret this result as: a characteristic of the top teams is that once they have a good pace for a few recent games in a row, then it is likely that they will perform well again. On the other hand, once we additionally incorporate earlier games as well, the prediction tends to become poorer.

The remaining seven teams (that is: the lower seven) tend to have a reasonably symmetric barplot, with the exception of Lotte Giants (7th ranked) that has a right-skewed shape. Such symmetries indicate that these teams don't have a particular value of $k$ such that the $k^{th}$ order Markov chain better models their outcomes compared to other orders. So either considering only the few recent games or taking more further past outcomes into account does not appear to have much difference in predicting the outcome. Perhaps one could think of this characteristic as: for these teams, the performance in recent games (regardless of which value $k$ takes out of $\{ 1, \cdots, 13 \}$) tend to not be influential to its performance today in the first place.

\section{Discussions and Future Work}
\noindent
Through our results we saw that the top three teams in the KBO league (Doosan, SK, and Hanwha) has a common characteristic that overall, lower values of $k$ tend to have the $k^{th}$ order Markov chain to better model their outcome. On the other hand, the remaining teams except Lotte has a reasonably symmetric shape in their barplot, meaning there is not really a particular value of $k$ that works considerably well compared to other values.

However, our analyses has limitations and thus potential future work that can improve the study. First, all of the interpretations were based on exploratory analyses. We plotted a barchart for each team, visually observing how well each value of $k$ did in terms of its $k^{th}$ order Markov chain predicting the match outcomes and thereby modeling the team's performance. We cannot make any formal conclusions at this point. To do so, we could utilize statistical tests on our data, but unfortunately, the size of our data is currently too small. For example, considering applying some kind of two-sample t-test for comparing the top-half teams and the bottom-half teams. We currently only have sample sizes of $n_{1} = 5$ and $n_{2} = 5$. One way of obtaining a larger sample would be to look for the observations in the past years of the KBO league: we go through each year's data, include the top-half ranked teams' sequences into group 1, and the other teams' sequences into group 2. This approach does have a risk that we have to assume that observations across different years for the same team are independent (by definition of the t-test). That is: we have to assume that the 2018 edition of Doosan Bears is independent of the 2017 edition of Doosan Bears, which, according to common sense, is not really a valid assumption to make. Another way would be to incorporate data of other leagues such as the MLB and the NPB since those leagues also have the characteristic that each team plays a game nearly everyday.

In addition, given the task to predict a team's game outcome, depending solely on the team's recent game results is perhaps an oversimplification of the task. Common sense tells us that there are other numerous factors that affect a team's performance in a game: e.g. the team's winning rate against the opponent in this season, statistics regarding the starting pitchers' of both teams, whether it's a home game or an away game, etc. So we could perhaps use a classical regression / classification model such as linear regression, support vector machines, deep neural networks, etc where we include those canonical features and additionally the result of the recent $k$ games as the predictors. Furthermore, if we want to stay with higher-order Markov chains but gain better modeling, we could consider using the higher-order multivariate Markov chains, where we are given $s$ separate categorical sequences that all have the same state space, instead of just one. The $k^{th}$ order multivariate Markov chain model says that the probability distribution (across the $m$ states) for the $j^{th}$ sequence at an arbitrary timestep depends on the probability distributions of all the $s$ sequences, including its own, at the recent $k$ timesteps. This model is also implemented in the the \textsf{markovchain} R package$^{4}$. In our study, this model can be utilized in such a way that given an arbitrary baseball game between two teams, the data sequence for both teams (so total $s=2$ sequences) are incorporated to better model the game result. That is: we consider the recent trend, or flow, of both of the two competing teams.




\end{document}